\documentclass[conference]{IEEEtran}

\pdfoutput=1
\usepackage{graphicx}
\usepackage{amssymb}
\usepackage{epstopdf}
\usepackage{float}
\usepackage{dblfloatfix}
\usepackage{amsmath}
\usepackage{algorithm}
\usepackage[noend]{algpseudocode}
\usepackage[justification=centering]{caption}
\newtheorem{definition}{Definition}
\newtheorem{example}{Example}
\newtheorem{remark}{Remark}

\begin{document}
%
\title{Index Coded PSK Modulation}
\author{Anjana~A.~Mahesh  and B~Sundar~Rajan,~\IEEEmembership{Fellow,~IEEE,}}
 
\maketitle

\begin{abstract}
 In this paper we consider noisy index coding problem over AWGN channel. We give an algorithm to map the index coded bits to appropriate sized  PSK symbols such that for the given index code, in general, the receiver with large amount of side information will gain in probability of error performance compared to the ones with lesser amount, depending upon the index code used. We call this the \textbf{PSK side information coding gain}. Also, we show that receivers with large amount of side information obtain this coding gain in addition to the bandwidth gain whereas receivers with lesser amount of side information trade off this coding gain with bandwidth gain. Moreover, in general, the difference between the best and worst performance among the receivers is shown to be proportional to the length of the index code employed. \footnote{The authors are with the Department of Electrical Communication Engineering, Indian Institute of Science, Bangalore-560012, India, Email: anjanaam@ece.iisc.ernet.in, bsrajan@ece.iisc.ernet.in}
\end{abstract}

\begin{IEEEkeywords}
Index coding, AWGN broadcast channel, M$-$PSK, PSK side information gain
\end{IEEEkeywords}

\IEEEpeerreviewmaketitle

\section{Introduction}
\label{sec:Intro}
\subsection{Preliminaries}
\label {sub:Prel}

\IEEEPARstart
{T}{he} noiseless index coding problem was first introduced by Birk and Kol \cite{ISCO} as an informed source coding problem over a broadcast channel. It involves a single source $\mathcal{S}$ that wishes to send $n$ messages from a set $X= \lbrace x_{1},x_{2},\ldots,x_{n} \rbrace $ to a set of $m$ receivers $\mathcal{R}=\lbrace R_{1},R_{2},\ldots,R_{m} \rbrace  $. The messages $\lbrace x_{1},x_{2},\ldots,x_{n} \rbrace $ belong to the finite field $\mathbb{F}_{2}$. A receiver $R_{i}$ $\in$ $\mathcal{R}$ is identified by $\lbrace \mathcal{W}_i , \mathcal{K}_i \rbrace$, where $\mathcal{W}_i\subseteq X$ is the set of messages demanded by the receiver $R_{i}$ and $\mathcal{K}_i\subsetneq X$ is the set of messages known to the receiver $R_{i}$ a priori. The index coding problem can be specified by $\left(X,\mathcal{R}\right)$.

\begin{definition}
An index code for the index coding problem $\left(X,\mathcal{R}\right)$ consists of\\$1)$An encoding function $f:\mathbb{F}_{2}^n \rightarrow \mathbb{F}_{2}^l$\\
$2)$A set of decoding functions $g_{1}, g_{2},\ldots,g_{m}$ such that, for a given input $\textbf{x} \in \mathbb{F}_{2}^n $, $g_{i}\left(f(\textbf{x}),\mathcal{X}_i\right) = \mathcal{W}_i, \ \forall i \in \lbrace 1, 2,\ldots,m \rbrace $.
\end{definition}
The optimal index code as defined in \cite{OMIC} is that index code which minimizes $l$, the number of binary transmissions.

An index code is said to be $linear$ if its encoding function is linear and $linearly$ $decodable$ if all its decoding functions are linear \cite{ICSI}. Further, \cite{ICSI}  establishes that for the class of index coding problems which can be represented using side information graphs, which were labeled 
later in \cite{OMIC} as $single$ $unicast$ index coding problems, the length of optimal linear index code is equal to the $minrank$ over $\mathbb{F}_{2}$ of the corresponding side information graph. This is extended in \cite{ECIC} to a general instance of index coding problem using minrank over $\mathbb{F}_q$ of the corresponding side information hypergraph.

In both \cite{ISCO} and \cite{ICSI}, noiseless binary channels are considered and hence the problem of index coding is formulated as a scheme to reduce the number of binary transmissions. This amounts to minimum bandwidth consumption, with binary transmission. We consider noisy index coding problems. We can reduce bandwidth further by using some M-ary modulation scheme. Hence we consider AWGN broadcast channel. A previous work which considered index codes over Gaussian broadcast channel is by L.Natarajan et al.\cite {IGBC}. Index codes based on multi-dimensional QAM constellations were proposed and a metric called $side\ information\ gain$ was introduced as a measure of efficiency with which the index codes utilizes receiver side information. However \cite {IGBC}  does not consider the index coding problem as originally defined in \cite{ISCO} and \cite{ICSI} as it does not minimize the number of transmissions. It always use $2^n$-point signal sets, where as we use signal sets of smaller sizes as well along with $2^n-$point signal sets for the same index coding problem.

\subsection{Our Contribution}
We consider index coding problems over AWGN broadcast channels. We find the length of the optimal linear index code of the given index coding problem by determining the minrank over $\mathbb{F}_{2}$ of the corresponding side information hypergraph, whenever possible, by brute force. Otherwise we use linear index codes of any length. Let the length of the index code be $N$. 
So a given input $\textbf{x} \in \mathbb{F}_{2}^n$ will result in a codeword $\textbf{c} \in \mathbb{F}_{2}^N $. Instead of using $N$ binary transmissions to broadcast the codeword $\textbf{c}$ as is done in noiseless index coding, we map the $N$-bit codeword to a $2^N$--PSK symbol with symbol energy equal to the total energy of the $N$ binary transmissions. By doing this, we get further gain in bandwidth, which we call the \textbf{PSK bandwidth gain}. In this paper, we propose an algorithm to map index coded bits into PSK symbols so that the receiver with maximum amount of side information gains in probability of error performance, followed by the receiver with next highest amount of side information and so on, which we term as the \textbf{PSK side\ information\ coding\ gain}(PSK-SICG). We show that there is a fundamental limit on the amount of side information a receiver should have so as to get PSK-SICG and that this limit is also influenced by the linear index code that we choose.  Also, we show that receivers with large amount of side information obtain this coding gain in addition to the bandwidth gain whereas receivers with lesser amount of side information trade off this coding gain with bandwidth gain. Moreover, in general, the difference between the best and worst performance among the receivers is shown to be proportional to the length of the index code employed. Exceptions to this general result are possible and Example \ref{ex5} is an instance. 

\subsection{Organization}
The rest of this paper is organized as follows. In Section \ref{sec:Model}, the index coding problem setting that we consider is formally defined with examples. The term PSK-SICG is formally defined. The fundamental limit on the amount of side information a receiver should have and its relation to the chosen index code in order to get PSK-SICG is also discussed. In Section \ref{sec:Algo}, we give an algorithm to map the index coded bits to a $2^N$--PSK symbol such that the receiver with maximum amount of side information sees maximum PSK-SICG. We go on to give examples with simulation results to support our claims in the subsequent Section \ref{sec:simu}. Finally the results are summarized in Section \ref{sec:conc}.


\section{Signal Model \& Preliminaries}
\label{sec:Model}
A general index coding problem can be converted into one where each receiver demands exactly one message, i.e.,$\left|\mathcal{W}_i \right| = 1,\ \forall i \in \lbrace 1, 2, \ldots, m\rbrace$. A receiver which demands more than one message, i.e., $\left|\mathcal{W}_i \right|>1$, can be considered as $\left|\mathcal{W}_i \right|$ equivalent receivers all having the same side information set  $\mathcal{K}_i$ and demanding a single message each. Since the same message can be demanded by multiple receivers, this gives $m \geq n$.

\begin{example}
\label{ex1} 
Let $m=n=$7. $\mathcal{W}_i = x_{i},\forall i\in \lbrace 1, 2,\ldots,7 \rbrace $. 
$\mathcal{K}_1 =\left\{2,3,4,5,6,7\right\},\ \mathcal{K}_2=\left\{1,3,4,5,7\right\},\ \mathcal{K}_3=\left\{1,4,6,7\right\},\mathcal{K}_4=\left\{2,5,6\right\},\mathcal{K}_5=\left\{1,2\right\},\ \mathcal{K}_6=\left\{3\right\},\mathcal{K}_7=\phi$.\\
The minrank over $\mathbb{F}_{2}$ of the side information graph corresponding to the above problem evaluates to $N$=4. 
An optimal linear index code is given by the encoding matrix,\\
\begin{center}
$L =\left[\begin{array}{cccc}
1 & 0 & 0 & 0\\
1 & 0 & 0 & 0\\
0 & 1 & 0 & 0\\
0 & 0 & 1 & 0\\
1 & 0 & 0 & 0\\
0 & 1 & 0 & 0\\
0 & 0 & 0 & 1
\end{array}\right]$. 
\end{center}
 The index coded bits are $\textbf{y}=\textbf{x}L$,  where,$$
  \textbf{y}= \left[y_1 \ y_2\ y_3\ y_4\right]=\left[x_1\ x_2\ \ldots x_7\right]L=\textbf{x}L$$
giving,
  \begin{align*}
    y_{1}&=x_{1}+x_{2}+x_{5}\\
    y_{2}&=x_{3}+x_{6} \\ 
    y_{3}&=x_{4} \\ 
    y_{4}&=x_{7} 
   \end{align*}
\end{example}

In the 4-fold BPSK index coding scheme we will transmit 4 BPSK symbols. In the scheme that we propose, we will map the index coded bits to the signal points of a 16-PSK constellation and transmit a single complex number thereby saving bandwidth. To keep energy per bit the same, the energy of the 16-PSK symbol transmitted will be equal to the total energy of the 4 transmissions in the 4-fold BPSK scheme.
\begin{example}
	\label{ex2} 
	Let $m=n=$ 6. $\mathcal{W}_1 = 1,\ \mathcal{W}_2=2, \ \mathcal{W}_3=3,\ \mathcal{W}_4=\lbrace1,4\rbrace, \ \mathcal{W}_5=5,\ \mathcal{W}_6=6$. 
	$\mathcal{K}_1 =\left\{2,3,4,5,6\right\},\ \mathcal{K}_2=\left\{1,3,4,5\right\},\ \mathcal{K}_3=\left\{1,2,4\right\},\mathcal{K}_4=\left\{2,3,6\right\},\mathcal{K}_5=\left\{3,4\right\},\ \mathcal{K}_6=\left\{5\right\}$.\\
	Convert $R_4$ which demands two messages into two receivers $R_4$ and $R_7$ with $\mathcal{W}_4=1,\ \mathcal{K}_4=\left\{2,3,6\right\},\ \mathcal{W}_7=4,\ \mathcal{K}_7=\left\{2,3,6\right\}$, which makes $m=7,\ n=6$
	The minrank over $\mathbb{F}_{2}$ of the side information hypergraph corresponding to the above problem evaluates to $N$=3. 
	An optimal linear index code is given by the encoding matrix,\\
	\begin{center}
		$L =\left[\begin{array}{ccc}
		1 & 0 & 0 \\
		1 & 0 & 0 \\
		1 & 0 & 0 \\
		0 & 1 & 0 \\
		0 & 0 & 1\\
		0 & 1 & 1 \\
		
		\end{array}\right]$.
	\end{center}
	
\noindent
Here,$$
	\textbf{y}= \left[y_1 \ y_2\ y_3\right]=\left[x_1\ x_2\ \ldots x_6\right]L=\textbf{x}L$$
and we have
	\begin{align*}
	y_{1}&=x_{1}+x_{2}+x_{3}\\
	y_{2}&=x_{4}+x_{6} \\ 
	y_{3}&=x_{5}+x_{6}
	\end{align*}

 Here, instead of transmitting 3 BPSK symbols, we will transmit a signal point of an 8-PSK constellation.
\end{example}
  In general, if for a particular index coding problem, the minrank over $\mathbb{F}_{2}$ of the corresponding side information hypergraph is $N$, then, instead of transmitting $N$ BPSK symbols we will transmit a single point from a $2^{N}$-PSK signal set with the energy of the $2^{N}$-PSK symbol being equal to $N$ times the energy of a BPSK symbol, i.e., equal to the total transmitted energy of the $N$ BPSK symbols. This scheme will give bandwidth gain in addition to the gain in bandwidth obtained by going from $n$ to $N$ BPSK transmissions utilizing side information. This extra gain is termed as PSK bandwidth gain.

\begin{definition}
	The term \textbf{PSK bandwidth gain} is defined as the factor by which the bandwidth required to transmit the index code is reduced, obtained while transmitting a $2^{N}$-PSK signal point instead of transmitting $N$ BPSK signal points. 
\end{definition}

For an index coding problem, there will be a reduction in required bandwidth by a factor of $N/2$, which will be obtained by all receivers.

With proper mapping of the index coded bits to PSK symbols, the algorithm for which is given in Section \ref{sec:Algo}, we will see that receivers with more amount of side information will get better performance in terms of probability of error, provided the side information available satisfies certain properties. This gain in error performance, which is solely due to the effective utilization of available side information by the proposed mapping scheme, is termed as PSK-SICG. Further, by sending the index coded bits as a $2^N$-PSK signal point, if a receiver gains in probability of error performance relative to a receiver in the N-fold BPSK transmission scheme, we say that the receiver gets PSK absolute coding gain(PSK-ACG).

\begin{definition}
	The term \textbf{PSK side information coding gain} is defined as the coding gain a receiver with side information gets relative to one with no side information, when the index code is transmitted as a signal point in a $2^{N}$-PSK.
\end{definition}

\begin{definition}
	The term \textbf{PSK Absolute Coding gain} is defined as the gain in probability of error performance obtained by any receiver in the $2^N$-PSK signal transmission scheme  relative to its performance in N-fold BPSK transmission scheme.
\end{definition}

For each of the receivers $R_{i},\ i\in \lbrace1,2,\ldots,m\rbrace$, define the set $S_{i}$ to be the set of all binary transmissions which $R_{i}$ knows a priori, i.e., $S_{i}= \lbrace y_{j}|y_{j}=\sum\limits_{k \in J }x_{k} ,\ J\subseteq \mathcal{K}_{i}\rbrace$. For example, in Example \ref{ex1}, $S_1= \{ y_2,y_3,y_4 \},$ $S_2= \{ y_3,y_4 \},$ $S_3= \{ y_3, y_4  \},$ $S_6= \{ y_3 \},$ and  $S_4=S_5=S_7= \phi.$

A receiver, $R_{i}$ will get PSK-SICG only if its available side information satisfies at least one of the following two conditions.

 \begin{align}
	n-\left|\mathcal{K}_i\right| < N \\
   \left|S_{i}\right|\geq 1
 \end{align}

The condition (2) above indicates how the PSK side information coding gain is influenced by the linear index code chosen. Different index codes for the same index coding problem will give different values of $\left|S_{i}\right|, \ i \in \lbrace 1,2, \ldots,m \rbrace$ and hence leading to possibly different PSK side information coding gains.

Consider the receiver $R_1$ in Example \ref{ex1}. It satisfies both the conditions with $n-\left|\mathcal{K}_1\right|  = 7-6 = 1 < 4$ and $\left|S_{1}\right| = 3 >1 $. For a particular message realization $(x_1, x_2, \dots, x_7)$, the only index coded bit $R_1$ does not know a priori is $y_1$. Hence there are only 2 possibilities for the received codeword at the receiver $R_1$. Hence it needs to decode to one of these 2 codewords, not to one of the 16 codewords that are possible had it not known any of $y_1,\ y_2,\ y_3,\ y_4$ a priori. Then we say that $R_1$ sees an effective codebook  of size 2. This reduction in the size of the effective codebook seen by the receiver $R_1$ is due to the presence of side information that satisfied condition (1) and (2) above. 

For a receiver to see an effective codebook of size $ < 2^N $, it is not necessary that the available side information should satisfy both the conditions. If at least one of the two conditions is satisfied, then that receiver will see an effective codebook of reduced size and hence will get PSK-SICG by proper mapping of index coded bits to $2^N$-PSK symbols. This can be seen from the following example.

\begin{example}
\label{ex3}
Let $m=n=$ 6. $\mathcal{W}_i = x_{i}, \ \forall i\in \lbrace 1, 2,\ldots,6 \rbrace $. 
$\mathcal{K}_1 =\left\{2,3,4,5,6\right\},\ \mathcal{K}_2=\left\{1,3,4,5\right\},\ \mathcal{K}_3=\left\{2,4,6\right\},\ \mathcal{K}_4=\left\{1,6\right\},\ \mathcal{K}_5=\left\{3\right\},\ \mathcal{K}_6=\phi$.\\
The minrank over $\mathbb{F}_{2}$ of the side information graph corresponding to the above problem evaluates to $N$=4. 
An optimal linear index code is given by the encoding matrix,\\

\begin{center}
	
	$L = \left[\begin{array}{cccc}
	1 & 0 & 0 & 0\\
	0 & 1 & 0 & 0\\
	0 & 1 & 0 & 0\\
	1 & 0 & 0 & 0\\
	0 & 0 & 1 & 0\\
	0 & 0 & 0 & 1
	\end{array}\right]$ \\
	
\end{center}

The index coded bits in this example are,
\begin{align*}
y_{1}&=x_{1}+x_{4}\\
y_{2}&=x_{2}+x_{3} \\ 
y_{3}&=x_{5} \\ 
y_{4}&=x_{6}.
\end{align*}
\end{example}\
Here, receiver $R_4$ does not satisfy condition (1) since $n-\left|\mathcal{K}_4\right|  = 6-2 = 4 = N$.
However, it will still see an effective codebook of size 8, since  $\left|S_{4}\right| = 1 $, and hence will get PSK-SICG by proper mapping of the codewords to 16-PSK signal points.
\begin{figure*}
	\includegraphics[scale=0.3]{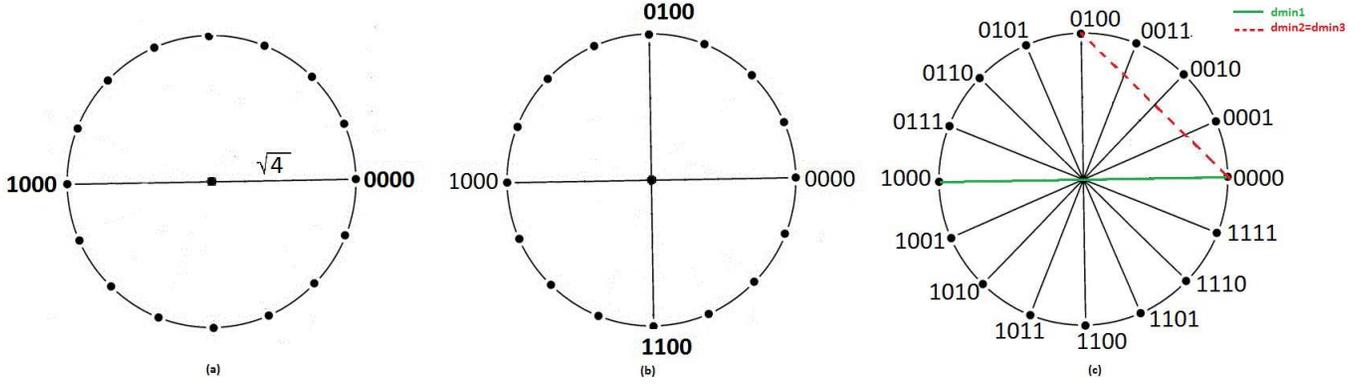}
	\caption{16-PSK Mapping for Example \ref{ex1}.}
	\label{fig1}
\end{figure*}
\begin{figure}[h]
	\includegraphics[scale=0.45]{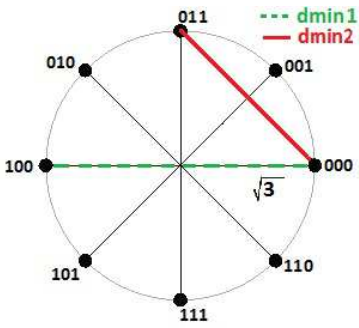}
	\caption{8-PSK Mapping for Example \ref{ex2}.}
	\label{fig2}
\end{figure}

\section{Algorithm}
\label{sec:Algo}
In this section we present the algorithm for labelling the appropriate sized PSK signal set. Let the minimum number of binary transmissions required $=$ minrank over $\mathbb{F}_{2}=N$ and the $N$ transmissions are labeled $Y= \lbrace y_{1},y_{2},\ldots,y_{N} \rbrace$, where each of $y_{i}$ is a linear combination of $\lbrace x_{1},x_{2},\ldots,x_{n} \rbrace$. If the minrank is not known then $N$ can be taken to be the length of any known linear index code. 

Let $\mathcal{C} = \lbrace \textbf{y} \in \mathbb{F}_2^N \ | \ \textbf{y} = \textbf{x}L ,\ \textbf{x} \in \mathbb{F}_2^n \rbrace$, where $L$ is the encoding matrix corresponding to the  linear index code chosen. Since $N \leq n $, $\mathcal{C} = \mathbb{F}_2^N$.

We have to consider the two conditions given by (1) and (2) listed in Section \ref{sec:Model}. 
Let $\eta_{i} \triangleq min \lbrace n-\left|\mathcal{K}_i\right|, N-\left|S_i\right|\rbrace$. The two conditions (1) and (2) are equivalent to the condition $\eta_{i} < N$. 

Order the receivers in the non-decreasing order of $\eta_{i}$.
WLOG, let $\lbrace R_{1},R_{2},\ldots,R_{m} \rbrace$ be such that 
\begin{align*}
\eta_{1} \leq \eta_2 \leq \ldots \leq \eta_m.
\end{align*}

Let $\mathcal{K}_i = \lbrace i_{1}, i_2, \ldots, i_{\left|\mathcal{K}_i\right|} \rbrace$ and $\mathcal{A}_i$ $\triangleq$ $\mathbb{F}_2^{\left|\mathcal{K}_i\right|}$, $i=1,2,\ldots,m$.
For any given realization of $( x_{i_1},x_{i_2}, \ldots, x_{i_{\left|\mathcal{K}_i\right|}} )$, the effective signal set seen by the receiver $R_i$ consists of $2^{\eta_i}$ points. Hence if $\eta_{i} \geq N $, then $d_{min}(R_i) \triangleq$ the minimum distance of the signal set seen by the receiver $R_i$, $i = 1,2,\ldots,m$, will not increase. $d_{min}(R_i)$ will remain equal to the minimum distance of the  corresponding $2^{N}$-- PSK. Thus for receiver $R_i$ to get PSK-SICG, $\eta_{i}$ should be less than $ N $.

The algorithm to map the index coded bits to PSK symbols is given in \textbf{Algorithm 1}.

		\begin{algorithm}
			\caption{Algorithm to map index coded bits to PSK symbols}\label{algo1}
			\begin{algorithmic}[1]
				
				\If {$\eta_1 \geq N $}, do an arbitrary order mapping and \textbf{exit}.
				\EndIf
				
				\State $i \gets 1$
				
				\If {all $2^N$ codewords have been mapped}, \textbf{exit}.
				\EndIf
				
				\State  Fix $( x_{i_1},x_{i_2}, \ldots, x_{i_{\left|\mathcal{K}_i\right|}} )=(a_{1},a_{2}, \ldots, a_{\left|\mathcal{K}_i\right|}) \in \mathcal{A}_i$ such that the set of codewords, $\mathcal{C}_i \subset \mathcal{C} $, obtained by running all possible combinations of $\lbrace x_{j}|\ j \notin \mathcal{K}_i\rbrace$ with $( x_{i_1},x_{i_2}, \ldots, x_{i_{\left|\mathcal{K}_i\right|}} )=(a_{1},a_{2}, \ldots, a_{\left|\mathcal{K}_i\right|})$ has maximum overlap with the codewords already mapped to PSK signal points.
				
				\If {all codewords in $\mathcal{C}_i$ have been mapped},
				\begin{itemize}
					\item $\mathcal{A}_i$=$\mathcal{A}_i \setminus \lbrace( x_{i_1},x_{i_2}, \ldots, x_{i_{\left|\mathcal{K}_i\right|}} )|( x_{i_1},x_{i_2}, \ldots, x_{i_{\left|\mathcal{K}_i\right|}} )$ together with all combinations of $\lbrace x_{j}|\ j \notin \mathcal{K}_i\rbrace$ will result in $\mathcal{C}_i\rbrace$.
					\item $i \gets i+1$
					\item \textbf{if} {$\eta_i \geq N$} \textbf{then},
					\begin{itemize}
					\item $i \gets 1$.
					\item goto \textbf{Step 3}
					\end{itemize} 
					\item \textbf{else}, goto \textbf{Step 3}
					
				\end{itemize}

				\Else
				\begin{itemize}
					\item Of the codewords in $\mathcal{C}_i$ which are yet to be mapped, pick any one and map it to a PSK signal point such that this point together with the signal points corresponding to already mapped codewords in $\mathcal{C}_i$, has the largest minimum distance possible. Clearly this minimum distance, $d_{min}(R_i)$ is such that $d_{min}$ of $ 2^{\eta_i}$-PSK $\geq d_{min}(R_i) \geq d_{min}$ of $ 2^{\eta_i +1}$-PSK.
					\item $i \gets 1$
					\item goto \textbf{Step 3}
				\end{itemize}
				\EndIf

			\end{algorithmic}
		\end{algorithm}

		\begin{remark}
			Note that the Algorithm \ref{algo1} above does not result in a unique mapping of index coded bits to $2^N$-PSK symbols. The mapping will change depending on the choice of $( x_{i_1},x_{i_2}, \ldots, x_{i_{\left|\mathcal{K}_i\right|}} )$ in each step. However, the performance of all the receivers obtained using any such mapping scheme resulting from the algorithm will be the same. Further, if $\eta_{i}= \eta_j$ for some $i \neq j$, depending on the ordering of $\eta_i$ done before starting the algorithm, $R_i$ and $R_j$ may give different performances in terms of probability of error.
		\end{remark}

\section{Simulation Results}
\label{sec:simu}

Consider the index coding problem in Example \ref{ex1} in Section \ref{sec:Model}.
Here, $\eta_1=1,\ \eta_2= \eta_3 =2$ and $\eta_i \geq 4, \  i \in \lbrace4,5,6,7\rbrace$.
Running the Algorithm \ref{algo1} in Section  \ref{sec:Algo}, suppose we fix $( x_2, x_3, x_4, x_5, x_6, x_7)= (000000)$,  we get $\mathcal{C}_{1}=\lbrace \lbrace0000\rbrace, \lbrace1000\rbrace \rbrace$. These codewords are mapped to diametrically opposite 16-PSK symbols as shown in Fig.  \ref{fig1}(a). Then, $\mathcal{C}_{2}$, which results in maximum overlap with $\lbrace \lbrace0000\rbrace, \lbrace1000\rbrace \rbrace$, is  $\lbrace \lbrace0000\rbrace,  \lbrace0100\rbrace, \lbrace1000\rbrace,\lbrace1100\rbrace \rbrace$. We consider $\lbrace0100\rbrace \in \mathcal{C}_2 \setminus \lbrace \lbrace0000 \rbrace, \lbrace1000\rbrace \rbrace$ and map it to a signal point such that these three codewords are at the best possible minimum distance. Now we go back to Step 3 with $i=1$ and find $C_1$ which has maximum overlap with the mapped codewords. Now $C_{1} =\lbrace \lbrace0100\rbrace, \lbrace1100\rbrace \rbrace$. Then we map $\lbrace 1100 \rbrace \in \mathcal{C}_1$ which is not already mapped, to a PSK signal point such that $C_{1} =\lbrace \lbrace0100\rbrace, \lbrace1100\rbrace \rbrace$ has the maximum possible minimum distance. This will result in the mapping  as shown in Fig.  \ref{fig1}(b). Continuing in this manner, we finally end up with the mapping shown in Fig. \ref{fig1}(c). We see that for such a mapping the $d_{min}^2(R_1)=(2\sqrt{(4)})^2=16$ and  $d_{min}^2(R_2)= d_{min}^2(R_3)=(\sqrt{2}\sqrt{4})^2=8$. 

\begin{figure*}
	
	\includegraphics[scale=0.4]{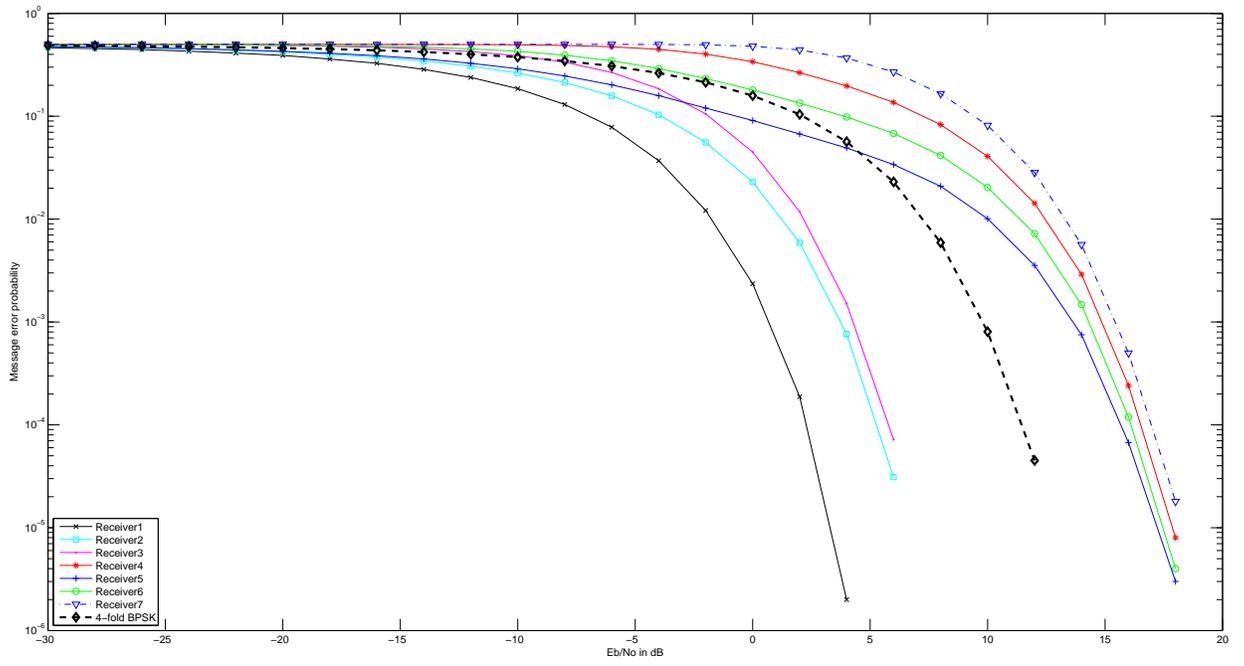}
	\caption{Simulation results for Example \ref{ex1}.}
	\label{fig3}
\end{figure*}

Simulation results for the above example is shown in Fig. \ref{fig3}. We see that the probability of message error plots corresponding to $R_{1}$ is well to the left of the plots of $R_{2}$ and $R_{3}$, which themselves are far to the left of other receivers as $R_{1},\ R_{2},\ R_{3}$ get PSK-SICG as defined in Section \ref{sec:Model}. Since $\left|S_{1}\right|>\left|S_{2}\right|=\left|S_{3}\right|, R_{1}$ gets the highest PSK-SICG. Further, since $\mathcal{K}_{4}, \ \mathcal{K}_{5},\ \mathcal{K}_{6}$ and $\mathcal{K}_{7}$ does not satisfy any of the two conditions required, they do not get PSK-SICG. The performance improvement gained by $R_1, R_2$ and $R_3$ over N-fold BPSK index code transmission can also be observed.

From the probability of message error plot, though it would seem that the receivers $R_{4}, R_{5}, R_{6}$ and $R_{7}$ lose out in probability of message error performance to the N-fold BPSK scheme, they are merely trading off coding gain for bandwidth gain as where the N-fold BPSK scheme for this example uses 4 real dimensions, the proposed scheme only uses 1 complex dimension, i.e., 2 real dimensions. Hence the receivers $R_{4},\ R_{5},\ R_{6}$ and $R_{7}$ get PSK bandwidth gain even though they do not get PSK-ACG whereas $R_{1}$, $R_{2}$ and $R_{3}$ get both PSK bandwidth gain and PSK-ACG. The amount of PSK-SICG, PSK bandwidth gain and PSK-ACG that each receiver gets is summarized in TABLE \ref{Table1}.
                                                                                                           
{\footnotesize
\begin{table}[h]
\renewcommand{\arraystretch}{2}
\begin{center}

\begin{tabular}{|c|c|c|c|c|c|c|c|}
	\hline
	 Parameter & $R_{1}$ & $R_{2}$ & $R_{3}$ & $R_{4}$ & $R_{5}$ & $R_{6}$ & $R_{7}$ \\
	 \hline 
	$d_{min_{PSK}}^2$ & 16 & 8 & 8 & 0.61 & 0.61 & 0.61 & 0.61  \\ 
	
	$d_{min_{binary}}^2$ & 4 & 4 & 4 & 4 & 4 & 4 & 4 \\ 
	
	PSK bandwidth gain & 2 & 2 & 2 & 2 & 2 & 2 & 2 \\
	
	PSK-SICG (in dB) & 14.19 & 11.19 & 11.19 & 0 & 0 & 0 & 0 \\
	
	PSK-ACG (in dB) & 6.02 & 3.01 & 3.01 & -8.16 & -8.16 & -8.16 & -8.16\\
	
	\hline
	
\end{tabular}

\caption \small { Table showing  PSK-SICG, PSK bandwidth gain and PSK-ACG for different receivers in Example \ref{ex1}.}

\label{Table1}	
	
\end{center}
\end{table}
}
\begin{table}[h]
	\renewcommand{\arraystretch}{2}
	\begin{center}
		
		\begin{tabular}{|c|c|c|c|c|c|c|}
			\hline
			Parameter & $R_{1}$ & $R_{2}$ & $R_{3}$ & $R_{4}$ & $R_{5}$ & $R_{6}$  \\
			\hline 
			$d_{min_{PSK}}^2$ & 12 & 6 & 1.76 & 1.76 & 1.76 & 1.76   \\ 
			
			$d_{min_{binary}}^2$ & 4 & 4 & 4 & 4 & 4 & 4 \\ 
			
			PSK bandwidth gain & 1.5 & 1.5 & 1.5 & 1.5 & 1.5 & 1.5  \\
			
			PSK-SICG (in dB) & 8.33 & 5.33 & 0 & 0 & 0 & 0  \\
			
			PSK-ACG (in dB) & 4.77 & 1.77 & -3.56 & -3.56 & -3.56 & -3.56 \\
			
			\hline
			
		\end{tabular}
		
		\caption \small { Table showing  PSK-SICG, PSK bandwidth gain and PSK-ACG for different receivers in Example \ref{ex2}.}
		
		\label{Table2}	
		
	\end{center}
\end{table}
\begin{figure*}
	
	\includegraphics[scale=0.4]{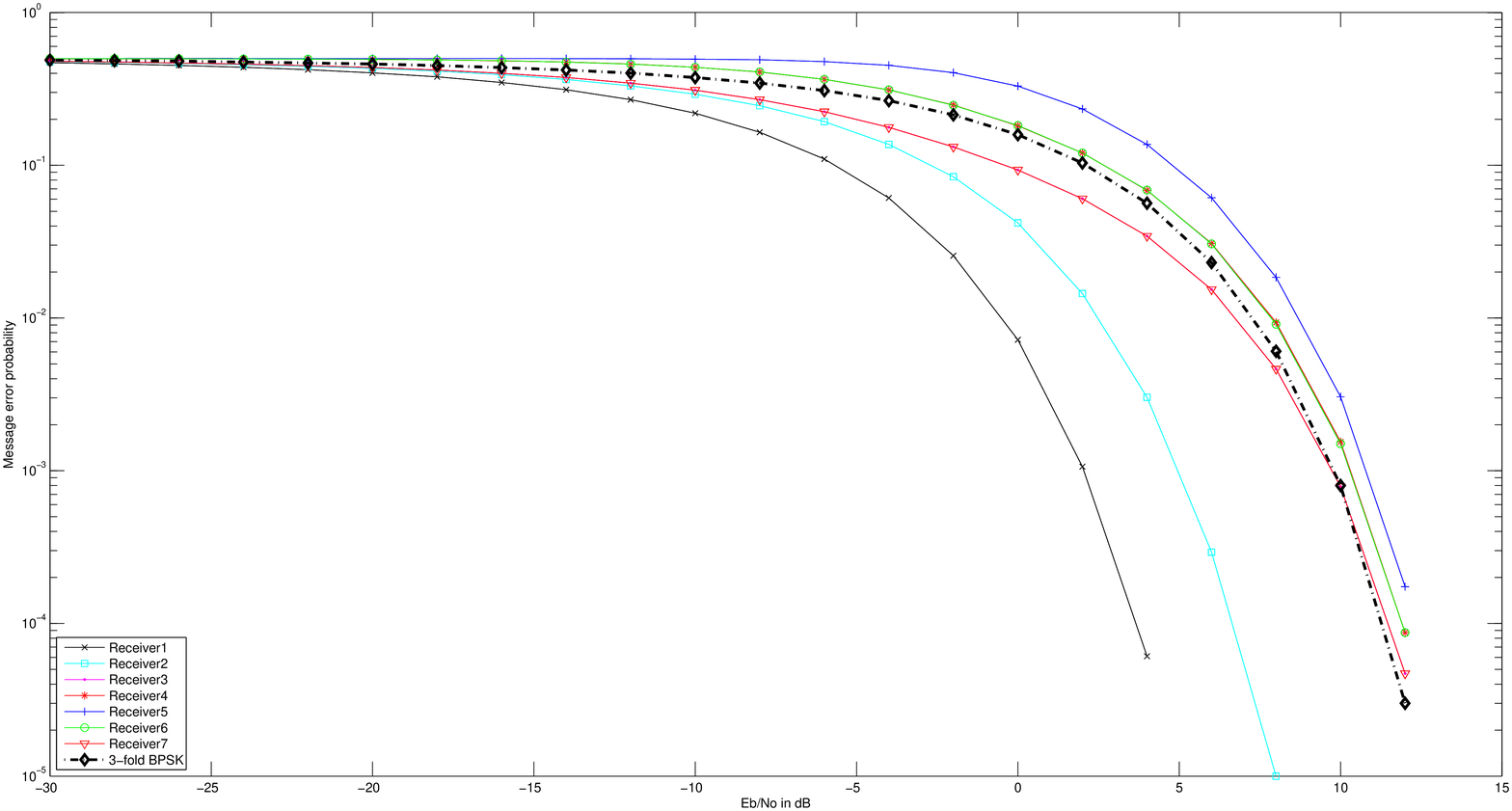}
	\caption{Simulation results for Example \ref{ex2}.}
	\label{fig4}
\end{figure*}
\begin{figure*}
	
	\includegraphics[scale=0.4]{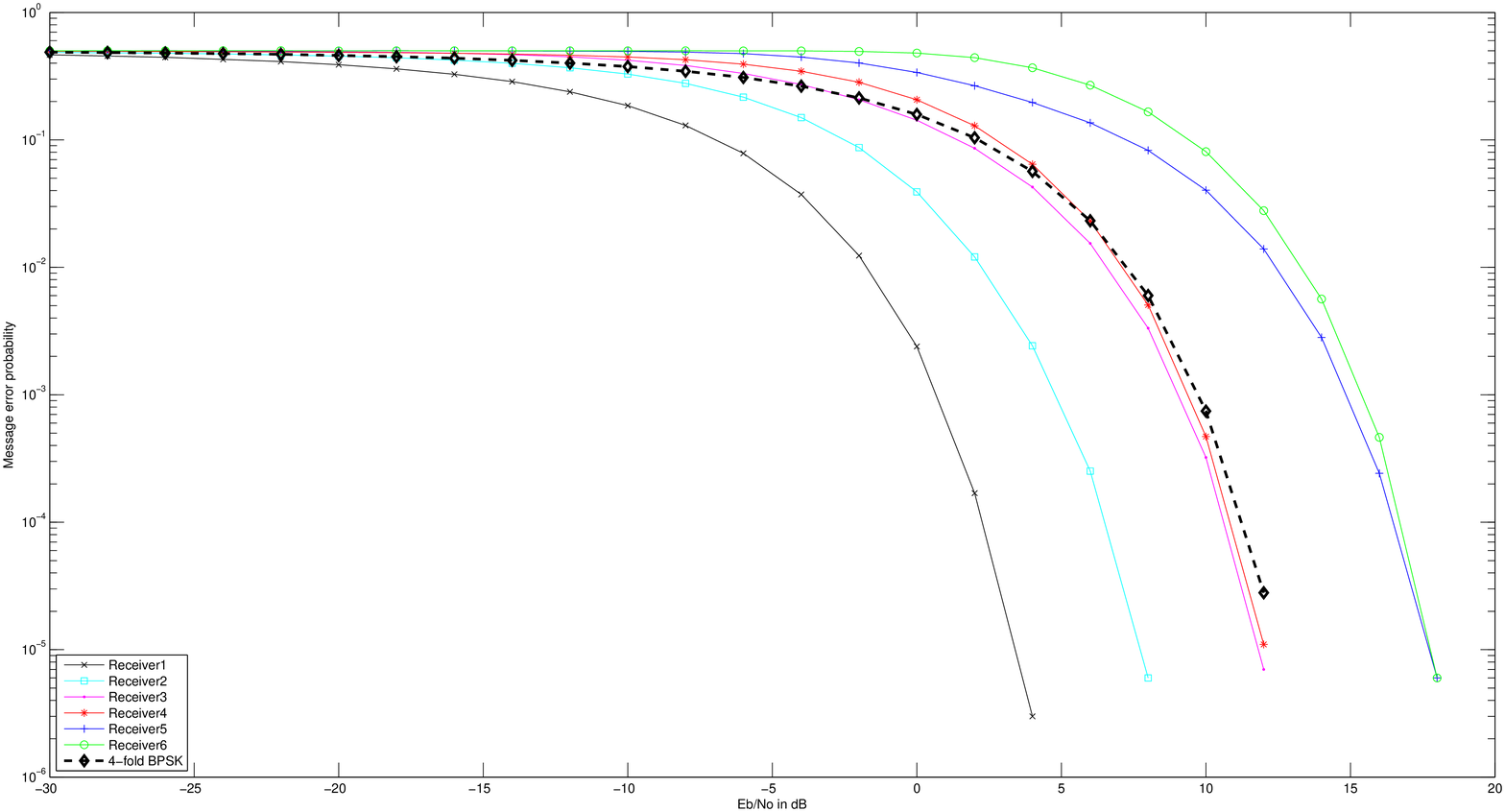}
	\caption{Simulation results for Example \ref{ex3}.}
	\label{fig5}
\end{figure*}
\begin{figure}
	\includegraphics[scale=0.42]{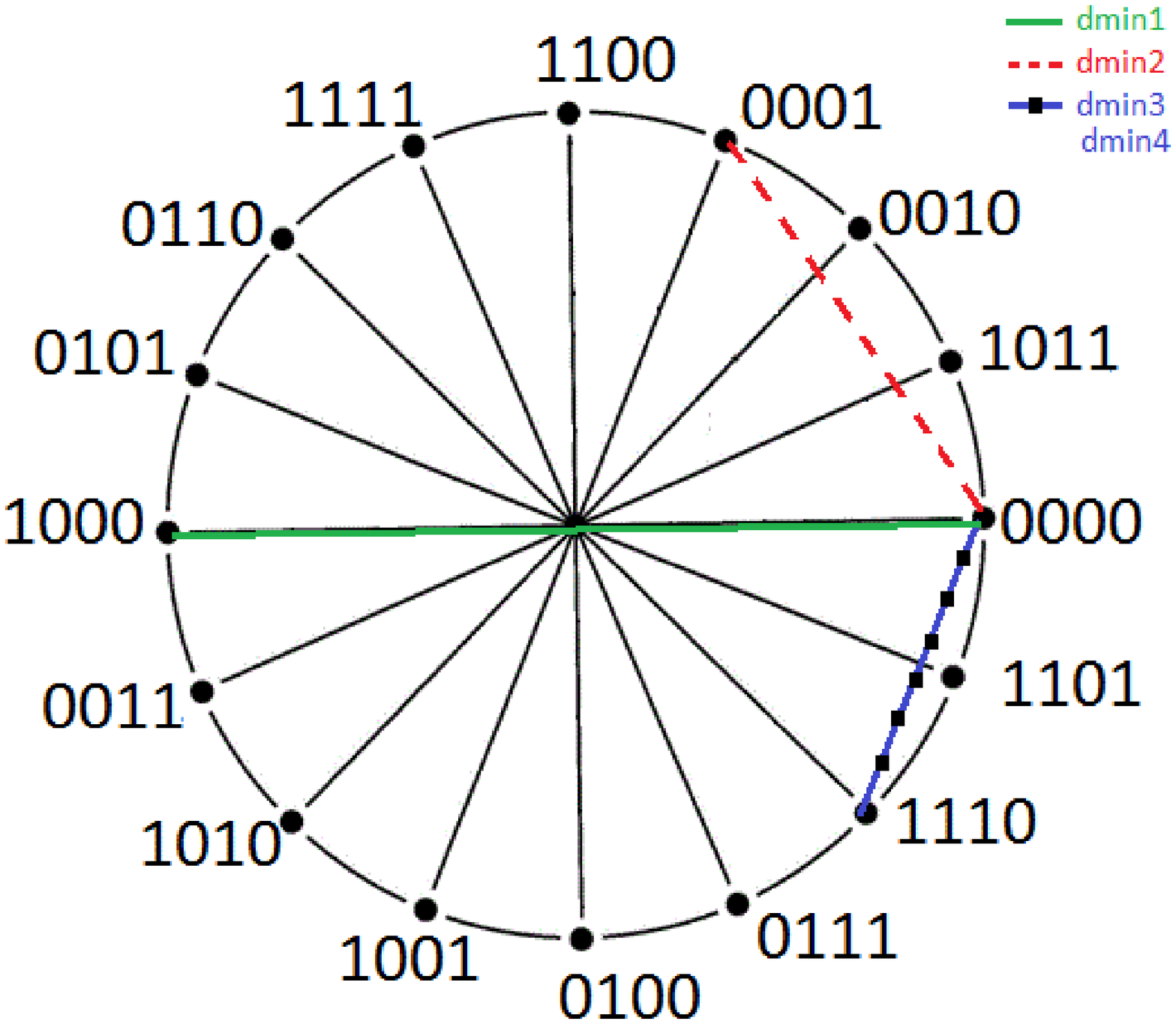}
	\caption{16-PSK Mapping for Example \ref{ex3}.}
	\label{fig6}
\end{figure}
Similarly a mapping for Example \ref{ex2} is shown in  Fig. \ref{fig2} and the simulation results are given in Fig. \ref{fig4}. We see that receivers $R_1$ and $R_2$ get PSK-SICG, PSK-ACG in addition to PSK bandwidth gain. All other receivers get PSK bandwidth gain and probability of error performance same as that of the N-fold BPSK scheme. The amount of PSK-SICG, PSK bandwidth gain and PSK-ACG that each receiver gets is summarized in TABLE \ref{Table2}.

Now consider Example \ref{ex3}. Here, suppose we fix  $( x_2, x_3, x_4, x_5, x_6)= (00000)$, we get  $\mathcal{C}_{1}=\lbrace \lbrace0000\rbrace, \lbrace1000\rbrace \rbrace$. After mapping these codewords, a subset of $\mathcal{C}$ which results in maximum overlap with already mapped codewords is  $\mathcal{C}_{2} = \lbrace \lbrace0000\rbrace,  \lbrace0001\rbrace, \lbrace0100\rbrace,\lbrace0101\rbrace \rbrace$.	We see that $\mathcal{C}_{1}\not \subseteq \mathcal{C}_{2}$, so codewords from $\mathcal{C}_{2}$ cannot be mapped to a 4-PSK signal set without disturbing the mapping of codewords of $\mathcal{C}_{1}$ already done. So we try to map them in such a way that the minimum distance,  $d_{min}(R_2) \geq d_{min}$ of  8-PSK. The algorithm gives a mapping which gives the best possible $d_{min}(R_2)$ keeping  $d_{min}(R_1) = d_{min}$ of 2-PSK. This mapping is shown in Fig. \ref{fig6}.

Simulation results for the above example is shown in Fig. \ref{fig5}. The receivers $R_{1}, R_{2}, R_{3}$ and $R_{4}$ get PSK-SICG. We see that the probability of message error plots corresponding to the N-fold BPSK binary transmission scheme lies near $R_{3}$ and $R_{4}$ showing better performances for receivers $R_{1}$ and $R_{2}$. Thus receivers $R_{1}$ and $R_{2}$ get PSK-ACG as well as PSK bandwidth gain over the N-fold BPSK scheme, $R_{3}$ and $R_{4}$ get the same performance as N-fold BPSK with additional bandwidth gain and $R_{5}$ and $R_{6}$ trade off bandwidth gain for coding gain. The amount of PSK-SICG, PSK bandwidth gain and PSK-ACG that each receiver gets is summarized in TABLE \ref{Table3}.

\begin{remark}
	Even though the minimum distance for the 4-fold BPSK transmissions is better than $d_{min}(R_3)$ and $d_{min}(R_4)$, as seen from TABLE \ref{Table3}, the probability of error plot for the 4-fold BPSK lies slightly to the right of the error plots for $R_3$ and $R_4$. This is because since the 4-fold BPSK scheme takes 2 times the bandwidth used by the 16-PSK scheme, the noise power = $N_o$(Bandwidth), where $N_o$ is the noise power spectral density, is 2 times more for the 4-fold BPSK. Therefore, the signal to noise power ratio for the 4-fold BPSK scheme is 2 times less than that for 16-PSK scheme, even though the transmitted signal power is the same for both the schemes. 
\end{remark}

\begin{table}[h]
	\renewcommand{\arraystretch}{2}
	\begin{center}
		
		\begin{tabular}{|c|c|c|c|c|c|c|}
			\hline
			Parameter & $R_{1}$ & $R_{2}$ & $R_{3}$ & $R_{4}$ & $R_{5}$ & $R_{6}$  \\
			\hline 
			$d_{min_{PSK}}^2$ & 16 & 4.94 & 2.34 & 2.34 & 0.61 & 0.61   \\ 
			
			$d_{min_{binary}}^2$ & 4 & 4 & 4 & 4 & 4 & 4 \\ 
			
			PSK bandwidth gain & 2 & 2 & 2 & 2 & 2 & 2  \\
			
			PSK-SICG (in dB) & 14.19 & 9.08 & 5.84 & 5.84 & 0 & 0  \\
			
			PSK-ACG (in dB) & 6.02 & 0.92 & -2.33 & -2.33 & -8.16 & -8.16 \\
			
			\hline
			
		\end{tabular}
		
		\caption \small { Table showing  PSK-SICG, PSK bandwidth gain and PSK-ACG for different receivers in Example \ref{ex3}.}
		
		\label{Table3}	
		
	\end{center}
\end{table}
\begin{example}
\label{ex4}
Let $m=n=$ 6. $\mathcal{W}_i = x_{i}, \ \forall i\in \lbrace 1, 2, \ldots, 6 \rbrace $. 
$\mathcal{K}_1 =\left\{2, 4, 5, 6\right\},\ \mathcal{K}_2=\left\{1, 3, 4, 5\right\},\ \mathcal{K}_3=\left\{2,4\right\},\ \mathcal{K}_4=\left\{1,3\right\},\ \mathcal{K}_5=\left\{2\right\},\ \mathcal{K}_6=\left\{1\right\}$.\\
For this problem, $N$=3. 
An optimal linear index code is given by the encoding matrix,\\

\begin{center}
	
$L = \left[\begin{array}{ccc}
1 & 0 & 0 \\
0 & 1 & 0 \\
0 & 0 & 1 \\
0 & 0 & 1 \\
0 & 1 & 0 \\
1 & 0 & 0 
\end{array}\right]$. \\
\end{center}

Here,\\
\begin{align*}
y_{1}=x_{1}+x_{6}\\
y_{2}=x_{2}+x_{5} \\ 
y_{3}=x_{3}+x_{4} \\ 
\end{align*}

\begin{figure}[h]
	\includegraphics[scale=0.38]{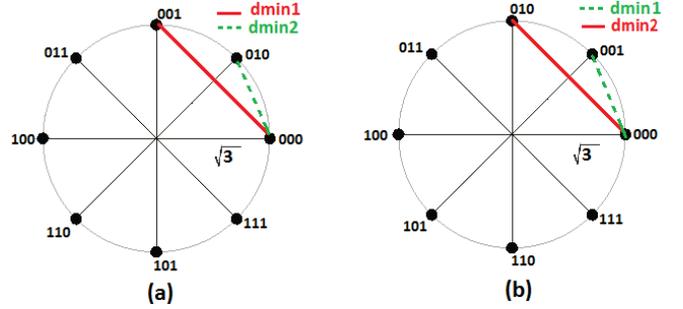}
	\caption{8-PSK Mappings for the 2 cases in Example \ref{ex4}.}
	\label{fig7}
\end{figure}

We see that $\left|\mathcal{K}_1\right| = \left|\mathcal{K}_2\right|$ and $\left|S_{1}\right| = \left|S_2\right|,\ \therefore \eta_1 = \eta_2$. Then, we can choose to prioritize $R_{1}$ or $R_{2}$ depending on the requirement. If we choose $R_{1}$, the resulting mapping is shown in Fig. \ref{fig7}(a) and if we choose $R_{2}$, then the mapping is shown in Fig. \ref{fig7}(b). Simulation results for this example with the mapping in Fig. \ref{fig7}(a) is shown in Fig. \ref{fig9}, where we can see that $R_{1}$ outperforms the other receivers. $R_{1}$ and $R_{2}$ get PSK-SICG as expected. They also get PSK-ACG. The other receivers have the same performance as the N-fold BPSK scheme. All 6 receivers get PSK bandwidth gain. The amount of PSK-SICG, PSK bandwidth gain and PSK-ACG that each receiver gets is summarized in TABLE \ref{Table4}.	

\begin{table}[h]
	\renewcommand{\arraystretch}{2}
	\begin{center}
		
		\begin{tabular}{|c|c|c|c|c|c|c|}
			\hline
			Parameter & $R_{1}$ & $R_{2}$ & $R_{3}$ & $R_{4}$ & $R_{5}$ & $R_{6}$  \\
			\hline 
			$d_{min_{PSK}}^2$ & 6 & 1.76 & 1.76 & 1.76 & 1.76 & 1.76   \\ 
			
			$d_{min_{binary}}^2$ & 4 & 4 & 4 & 4 & 4 & 4 \\ 
			
			PSK bandwidth gain & 1.5 & 1.5 & 1.5 & 1.5 & 1.5 & 1.5  \\
			
			PSK-SICG (in dB) & 5.33 & 0 & 0 & 0 & 0 & 0  \\
			
			PSK-ACG (in dB) & 1.77 & -3.56 & -3.56 & -3.56 & -3.56 & -3.56 \\
			
			\hline
			
		\end{tabular}
		
		\caption \small { Table showing  PSK-SICG, PSK bandwidth gain and PSK-ACG for different receivers for case (a) in Example \ref{ex4}.}
		
		\label{Table4}	
		
	\end{center}
\end{table}
\end{example}

Here, even though $d_{min}(R_2)=d_{min}(R_3)=d_{min}(R_4)=d_{min}(R_5)=d_{min}(R_6)$, the probability of error plot of $R_2$ is well to the left of the error plots of $R_3$, $R_4$, $R_5$ and $R_6$. This is because the distance distribution seen by $R_2$ is different from the distance distribution seen by the other receivers, as shown in TABLE \ref{Table5}, where, $d_{min_1}$ gives the minimum pairwise distance, $d_{min_2}$ gives the second least pairwise distance and so on.

\begin{table}
	\renewcommand{\arraystretch}{2}
	\begin{center}
		
		\begin{tabular}{|c|c|c|c|c|c|c|}
			\hline
			Parameter & $R_1$ & $R_2$ & $R_3$ & $R_4$ & $R_5$ & $R_6$ \\
			\hline
			Effective signal set seen & 4 pt & 4 pt & 8 pt & 8 pt & 8 pt & 8 pt \\
			$d_{min_1}$ & 6 & 1.76 & 1.76 & 1.76 & 1.76 & 1.76 \\
			No. of pairs & 4 & 4 & 8 & 8 & 8 & 8\\
			$d_{min_2}$ & 12 & 10.24 & 6 & 6 & 6 & 6 \\
			No. of pairs & 2 & 2 & 8 & 8 & 8 & 8 \\
			$d_{min_3}$ & -- & 12 & 10.24 & 10.24 & 10.24 & 10.24\\
			No. of pairs & 0 & 2 & 8 & 8 & 8 & 8 \\
			$d_{min_4}$ & -- & -- & 12 & 12 & 12 & 12 \\
			No. of pairs & 0 & 0 & 4 & 4 & 4 & 4 \\
			\hline			
		\end{tabular}
		\caption \small {Table showing the pair-wise distance distribution for the receivers in Example \ref{ex4}.}
		\label{Table5}
	\end{center}	
\end{table}


\begin{remark}
For the class of index coding problems, called single unicast single uniprior in \cite{OMIC}, $\left|\mathcal{S}_i\right| = 0, \ \forall i \in \lbrace 1,2, \ldots, m \rbrace$. Therefore, no receiver will get PSK-SICG.
\end{remark}

\subsection{$2^N$-PSK to $2^n$-PSK}
\label{sec6}
In this subsection we consider an example for which we show the performance for all the lengths from the minimum length $N$ to the maximum possible value of $n.$

Consider the following example.
%
\begin{example}
	\label{ex5}
	Let $m=n=$ 4. $\mathcal{W}_i = x_{i},\ \forall i\in \lbrace 1, 2, \ldots, 4 \rbrace $. 
	$\mathcal{K}_1 =\left\{2, 3, 4\right\},\ \mathcal{K}_2=\left\{1, 3\right\},\ \mathcal{K}_3=\left\{1,4\right\},\ \mathcal{K}_4=\left\{2\right\}$. \\
	For this problem, $N$=2. 
	An optimal linear index code is given by the encoding matrix,\\
	
	\begin{center}
		
		$L_1 = \left[\begin{array}{ccc}
		1 & 0  \\
		1 & 1  \\
		1 & 0  \\
		0 & 1  
	
		\end{array}\right]$.\\
	\end{center}
	
	Here,\\
	\begin{align*}
	y_{1}&=x_{1}+x_{2}+x_{3}\\
	y_{2}&=x_{2}+x_{4} \\ 
	\end{align*}
	The corresponding 4-PSK mapping is given in Fig. \ref{fig8}(a). 
	
	\begin{figure}[h]
		\includegraphics[scale=0.5]{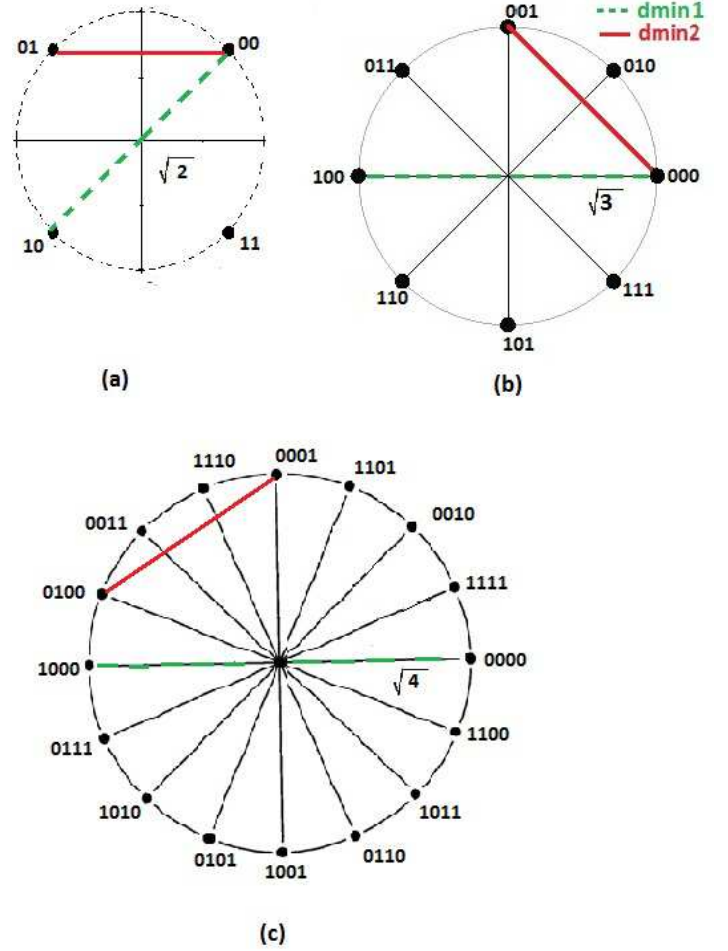}
		\caption{4-PSK, 8-PSK and 16-PSK Mappings for  Example \ref{ex5}.}
		\label{fig8}
	\end{figure}

	\begin{figure*}
		\includegraphics[scale=0.4]{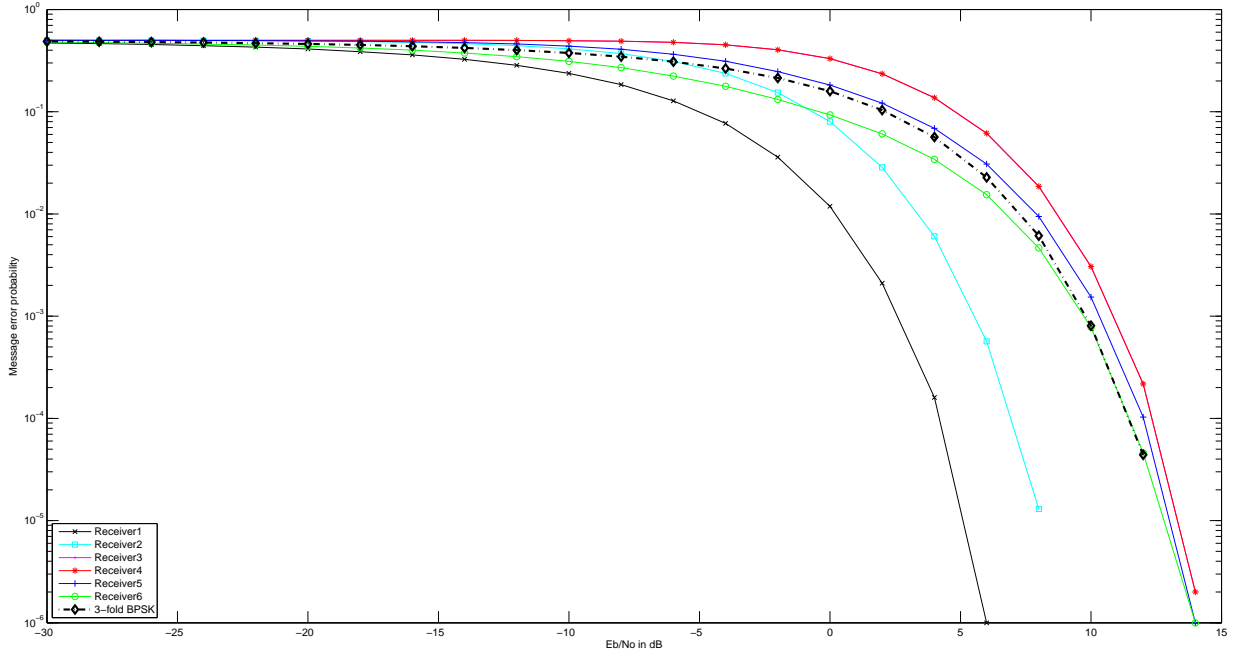}
		\caption{Simulation results for Example \ref{ex4}.}
		\label{fig9}
	\end{figure*}
	
	Now assuming  that we did not know the minrank for the above problem and chose $N=3$. Then an encoding matrix is, 
	\begin{center}
		
		$L_2 = \left[\begin{array}{ccc}
		1 & 0 & 0 \\
		1 & 0 & 0 \\
		0 & 1 & 0 \\
		0 & 0 & 1 \\
		
		\end{array}\right]$, \\
	\end{center}
	
	with,\\
	\begin{align*}
	y_{1}&=x_{1}+x_{2}\\
	y_{2}&=x_{3} \\ 
	y_{3}&=x_{4}\\ 
	\end{align*}
	
	and an 8-PSK mapping which gives the best possible PSK-SICGs for the different receivers is shown in Fig. \ref{fig8}(b).

	Now, compare the above two cases with the case where the 4 messages are transmitted as they are, i.e., $ \left[y_1 \ y_2\ y_3\ y_4\right]=\left[x_1\ x_2\ x_3\ x_4\right]$. A 16-PSK mapping which gives the maximum possible PSK-SICG is shown in Fig. \ref{fig8}(c).
	
	\begin{figure*}
		\includegraphics[scale=0.4]{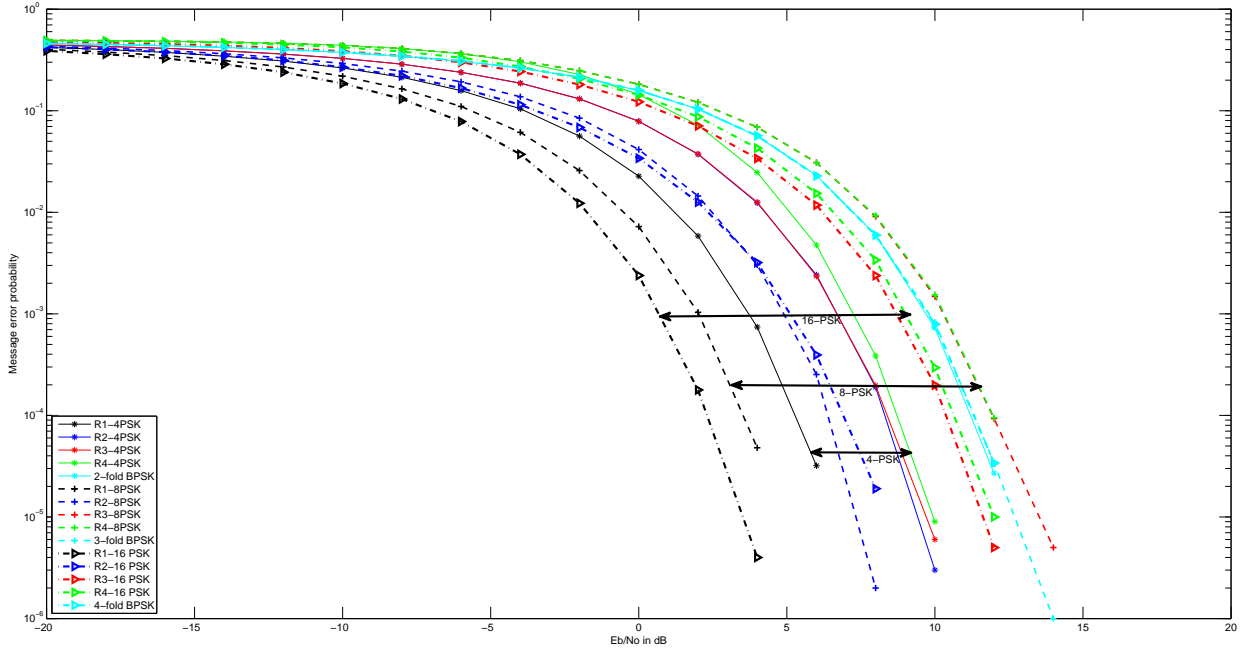}
		\caption{Simulation results for Example \ref{ex5}.}
		\label{fig10}
	\end{figure*}

	From the simulation results shown in Fig \ref{fig10}, we see that the performance of the best receiver, i.e., $R_1$, improves as we go from $N$ to $n$. However, the gap between the best performing receiver and worst performing receiver widens as we go from N to n. This is always the case if the worst performing receiver has no side information, as illustrated in the following example. However, if the worst performing receiver knows at least one message a priori, whether or not the gap widens monotonically depends on the mapping scheme used, as is the case with this example. The reason for the difference in performance seen by different receivers is that they see different minimum distances, which is summarized in TABLE \ref{Table6}, for 4-PSK, 8-PSK and 16-PSK.
	
	\begin{table}[h]
		\renewcommand{\arraystretch}{2}
		\begin{center}
			
			\begin{tabular}{|c|c|c|c|c|}
				\hline
				Parameter & $R_{1}$ & $R_{2}$ & $R_{3}$ & $R_{4}$  \\
				\hline 
				$d_{min_{\ 4-PSK}}^2$ & 8 & 4 & 4 & 4   \\ 
				
				$d_{min_{\ 8-PSK}}^2$ & 12 & 6 & 1.76 & 1.76   \\ 
					
				$d_{min_{\ 16-PSK}}^2$ & 16 & 4.94 & 2.34 & 2.34   \\ 
				
				$d_{min_{binary}}^2$ & 4 & 4 & 4 & 4  \\ 
							
				\hline
				
			\end{tabular}
			
			\caption \small { Table showing  the minimum distances seen by different receivers for 4-PSK, 8-PSK and 16-PSK in  Example \ref{ex5}.}
			
			\label{Table6}	
			
		\end{center}
	\end{table}
	
	\end{example}
\begin{figure*}
\begin{center}		
\includegraphics[scale=0.4]{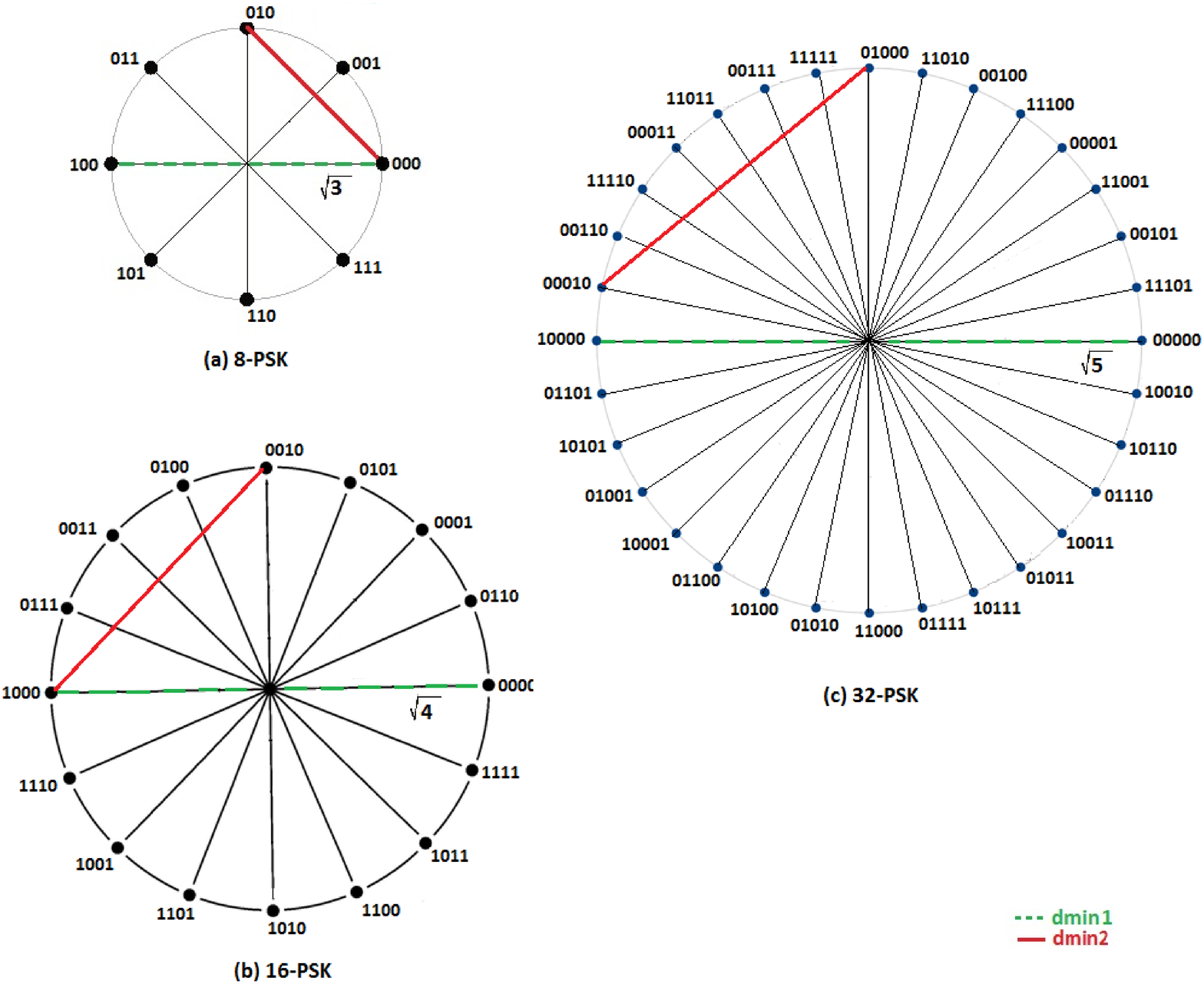}
		\caption{8-PSK, 16-PSK and 32-PSK Mappings for  Example \ref{ex6}.}
		\label{fig11}
\end{center}
%
\begin{center}	
	\includegraphics[scale=0.4]{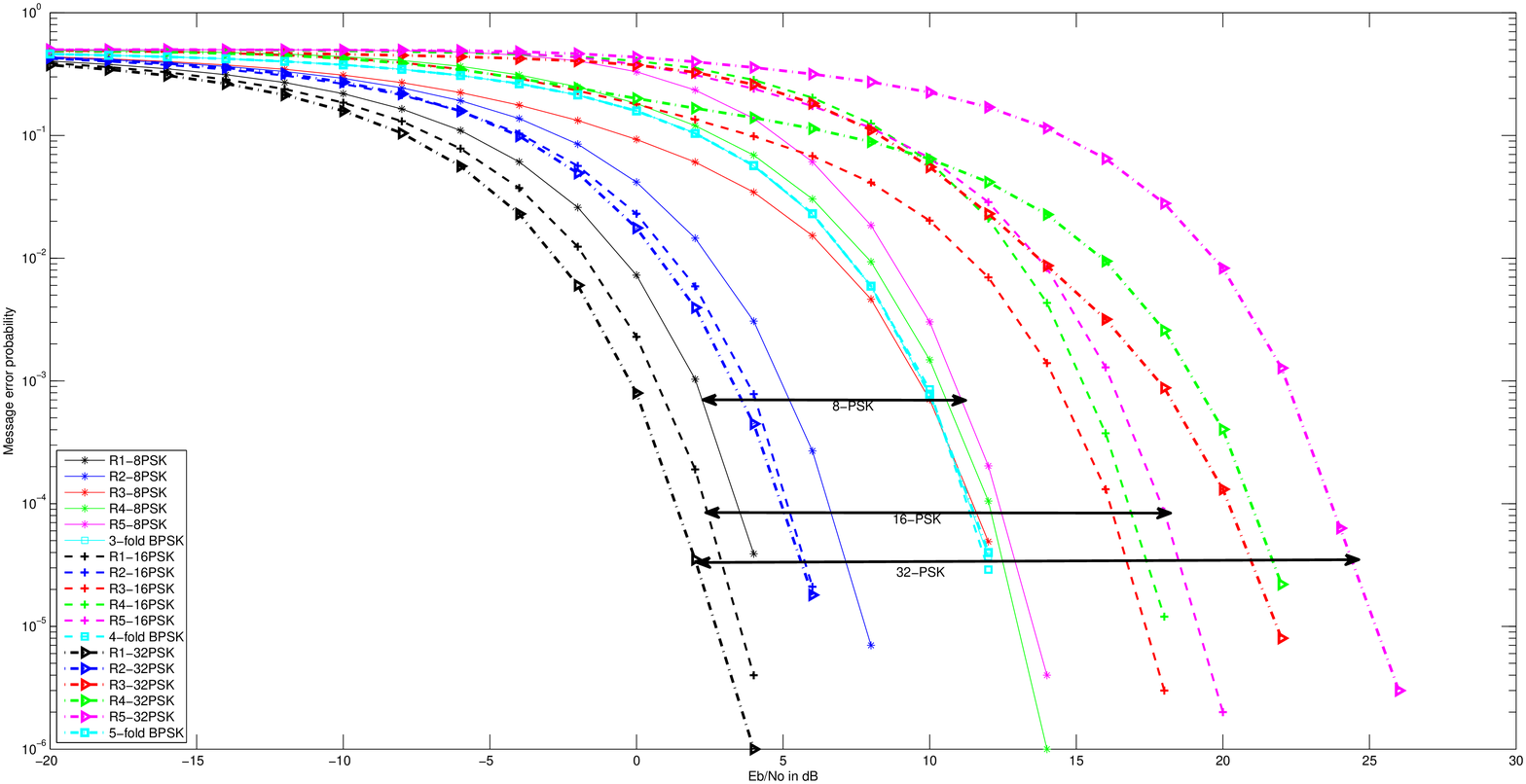}
		\caption{Simulation results for Example \ref{ex6}.}
		\label{fig12}
\end{center}	
\end{figure*}

	\begin{example}
	\label{ex6}
		Let $m=n=5.\ \mathcal{W}_i = \lbrace x_i \rbrace, \ \forall i \in \lbrace 1, 2, 3, 4, 5 \rbrace.\ \mathcal{K}_1=\lbrace2,3,4,5\rbrace, \ \mathcal{K}_2=\lbrace1,3,5\rbrace, \ \mathcal{K}_3=\lbrace1,4\rbrace, \ \mathcal{K}_4= \lbrace2\rbrace, \ \mathcal{K}_5 = \phi$.
		
		For this problem, minrank, $N$ = 3. An optimal linear index code is given by 
		\begin{align*}
		L_1 = \left[\begin{array}{ccc}
		1 & 0 & 0\\
		1 & 1 & 0\\
		1 & 0 & 0\\
		0 & 1 & 0\\
		0 & 0 & 1
		\end{array}\right],
		\end{align*}
		with the index coded bits being
$$ y_{1}=x_{1}+x_{2}+x_{3};~~ y_{2}=x_{2}+x_{4}; ~~ y_{3}=x_{5}. 
$$
%
%
		Now, we consider an index code of length $N+1=4$. The corresponding encoding matrix is 
		\begin{align*}
		L_2 = \left[\begin{array}{cccc}
		1 & 0 & 0 & 0\\
		1 & 0 & 0 & 0\\
		0 & 1 & 0 & 0\\
		0 & 0 & 1 & 0\\
		0 & 0 & 0 & 1
		\end{array}\right]
		\end{align*}
		and the index coded bits are
$$ y_{1}=x_{1}+x_{2};~~ y_{2}=x_{3}; y_{3}=x_{4}; y_{4}=x_{5}.
  $$
		We compare these with the case where we send the messages as they are, i.e., 
		\begin{align*}
		L_3= I_5, 
		\end{align*}
		where $I_5$ denotes the $5X5$ identity matrix.

	The PSK mappings which give performance advantage to receivers satisfying conditions (1) and (2) given in Section \ref{sec:Model} for the three different cases considered are given in Fig. \ref{fig11}(a), (b) and (c) respectively.
	
	\begin{table}[h]
		\renewcommand{\arraystretch}{2}
		\begin{center}
			
			\begin{tabular}{|c|c|c|c|c|c|}
				\hline
				Parameter & $R_{1}$ & $R_{2}$ & $R_{3}$ & $R_{4}$ & $R_{5}$ \\
				\hline 
				$d_{min_{\ 8-PSK}}^2$ & 12 & 6 & 1.76 & 1.76 & 1.76 \\ 
				
				$d_{min_{\ 16-PSK}}^2$ & 16 & 8 & 0.61 & 0.61 & 0.61   \\ 
				
				$d_{min_{\ 32-PSK}}^2$ & 20 & 8.05 & 0.19 & 0.19 & 0.19  \\ 
				
				$d_{min_{binary}}^2$ & 4 & 4 & 4 & 4 & 4 \\ 
				
				\hline
				
			\end{tabular}
			
			\caption \small { Table showing  the minimum distances seen by different receivers for 8-PSK, 16-PSK and 32-PSK in  Example \ref{ex6}.}
			
			\label{Table7}	
			
		\end{center}
	\end{table}
    \end{example}
	The simulation results for the above example are shown in Fig. \ref{fig12}.	From Fig. \ref{fig12}, we can see that the gap between $R_1$ and $R_6$ widens monotonically as we move from $N$ to $n$. However the best performing receiver's, i.e., $R_1$'s  performance improves as we go from $N$ to $n$. The minimum distance seen by different receivers for the 3 cases considered, namely, 8-PSK, 16-PSK and 32-PSK, are listed in TABLE \ref{Table7}.
	
\section{Conclusion}
\label{sec:conc}
The mapping and 2-D transmission scheme proposed in this paper is applicable to any index coding problem setting. In a practical scenario, we can use this mapping scheme to prioritize those customers who are willing to pay more, provided their side information satisfies the condition mentioned in Section \ref{sec:Model}. Further, the mapping scheme depends on the index code, i.e., the encoding matrix, $L$, chosen, since $L$ determines $\left|S_{i}\right|, \forall i \in \lbrace1, 2, \ldots, m\rbrace$. So we can even choose an $L$ matrix such that it favors our chosen customer, provided $L$ satisfies the condition that all users use the minimum possible number of binary transmissions to decode their required messages. 

\end{document}